\documentclass[a4paper,11pt]{article}
\usepackage{jinstpub}

\usepackage{subcaption}
\usepackage{here}
\usepackage[dvipsnames]{xcolor}
\usepackage[T1]{fontenc}
\usepackage{tgcursor}
\usepackage{verbatim}
\usepackage{graphbox}

\title{Electronics integration for the GE2/1 and ME0 GEM detector systems for the CMS Phase-2 Muon System Upgrade}

\author[a,1]{S.~Butalla,\note{Corresponding author.}}
\author[a]{M.~Hohlmann,}
\author{\\on behalf of the CMS Muon Group}

\affiliation[a]{Florida Institute of Technology, Melbourne, FL, USA}

\emailAdd{stephen.butalla@cern.ch}

\abstract{The Large Hadron Collider is currently undergoing its High Luminosity upgrade, which is set to increase the instantaneous luminosity by about a factor of five. Consequently, the Compact Muon Solenoid experiment is upgrading its muon spectrometer to cope with the increased muon flux in the forward region. The GE2/1 triple-Gas Electron Multiplier detector, which has recently entered the mass production phase, and the ME0 triple-GEM detector system, which is in the late prototyping phase, are undergoing electronics integration. These proceedings briefly discuss the frontend electronics for the GE2/1 and ME0 detector systems, the electronics integration testing process, and the future plans for the frontend electronics of these two detector systems by the CMS GEM Collaboration.}


\begin{document}
\maketitle
\flushbottom

\section{Introduction}\label{sec:int}
With the projected five-fold increase in instantaneous luminosity resulting from the High Luminosity upgrade of the Large Hadron Collider, the Compact Muon Solenoid (CMS) experiment is currently upgrading its muon spectrometer \cite{MuonTDR}. Two triple-Gas Electron Multiplier (GEM) detector systems -- the GE2/1, currently in the early mass-production phase, and the ME0, currently in the prototyping phase -- are undergoing frontend electronics integration.
%(see figure \ref{fig:endcap}).
%Electronics integration efforts for the GE2/1 are carried out at CERN, Rice University, Texas A\&M, and Florida Tech, and for the ME0 at CERN, UCLA, and Florida Tech.
These proceedings discuss the current status of the electronics integration effort on full-size GE2/1 and ME0 chamber prototypes by the CMS GEM Collaboration and the future prospects for the frontend electronics readout systems.

%\begin{figure}[!htp] 
%\centering
%\includegraphics[width=0.45\textwidth]{cmsQuadrant_annotated_TWEPP}
%\caption{\label{fig:endcap} A quarter of the $R$-$z$ cross-section of the upgraded CMS detector with the GE1/1 (first muon station), GE2/1 (second muon station), and ME0 (behind the high granularity calorimeter, HGCAL) detector systems. Adapted from \cite{MuonTDR}}
%\end{figure}

\section{The GE2/1 and ME0 detector systems}
\begin{figure}[!hp]
\centering
\includegraphics[width=0.23\linewidth]{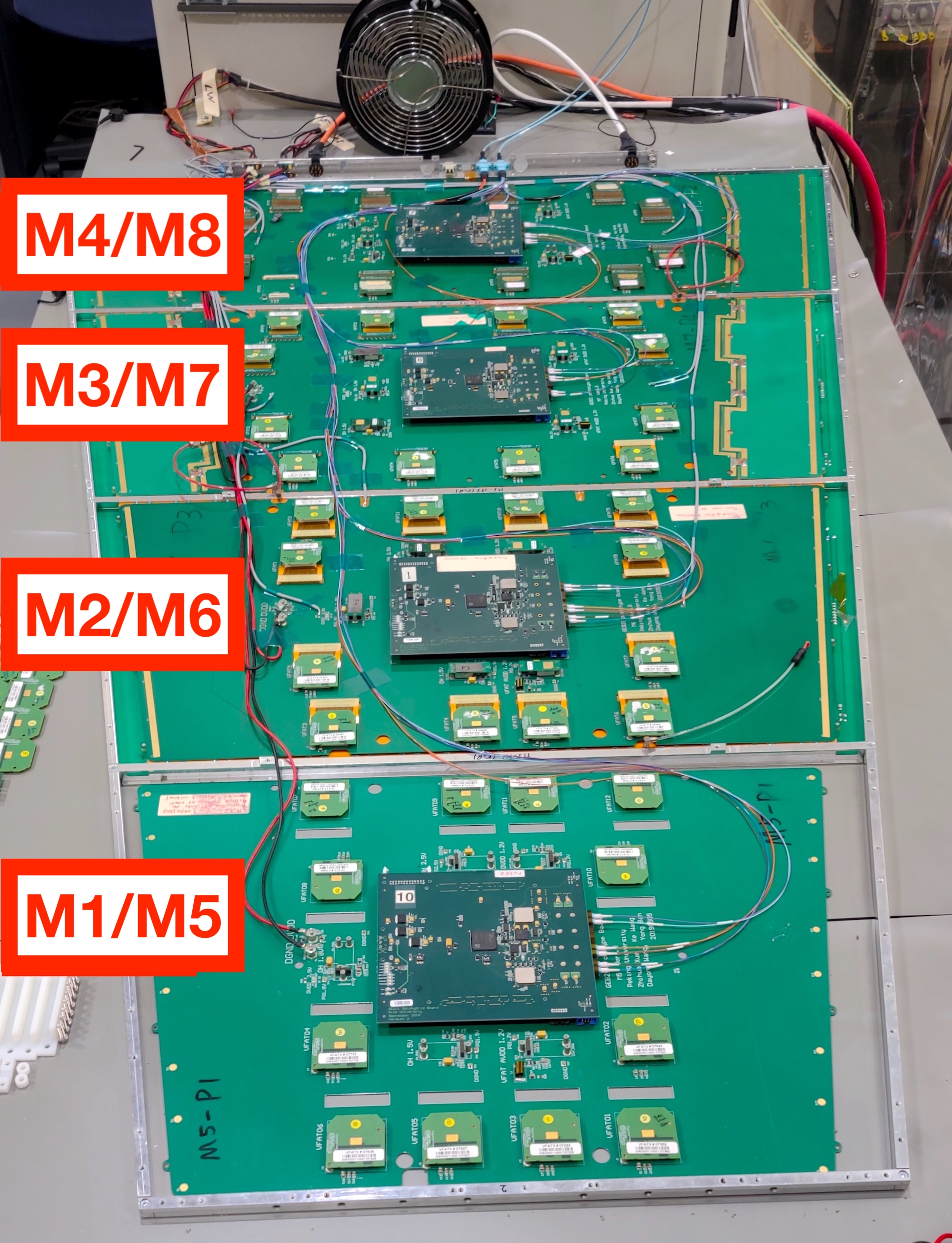}\hspace{0.4cm}\includegraphics[width=0.25\linewidth]{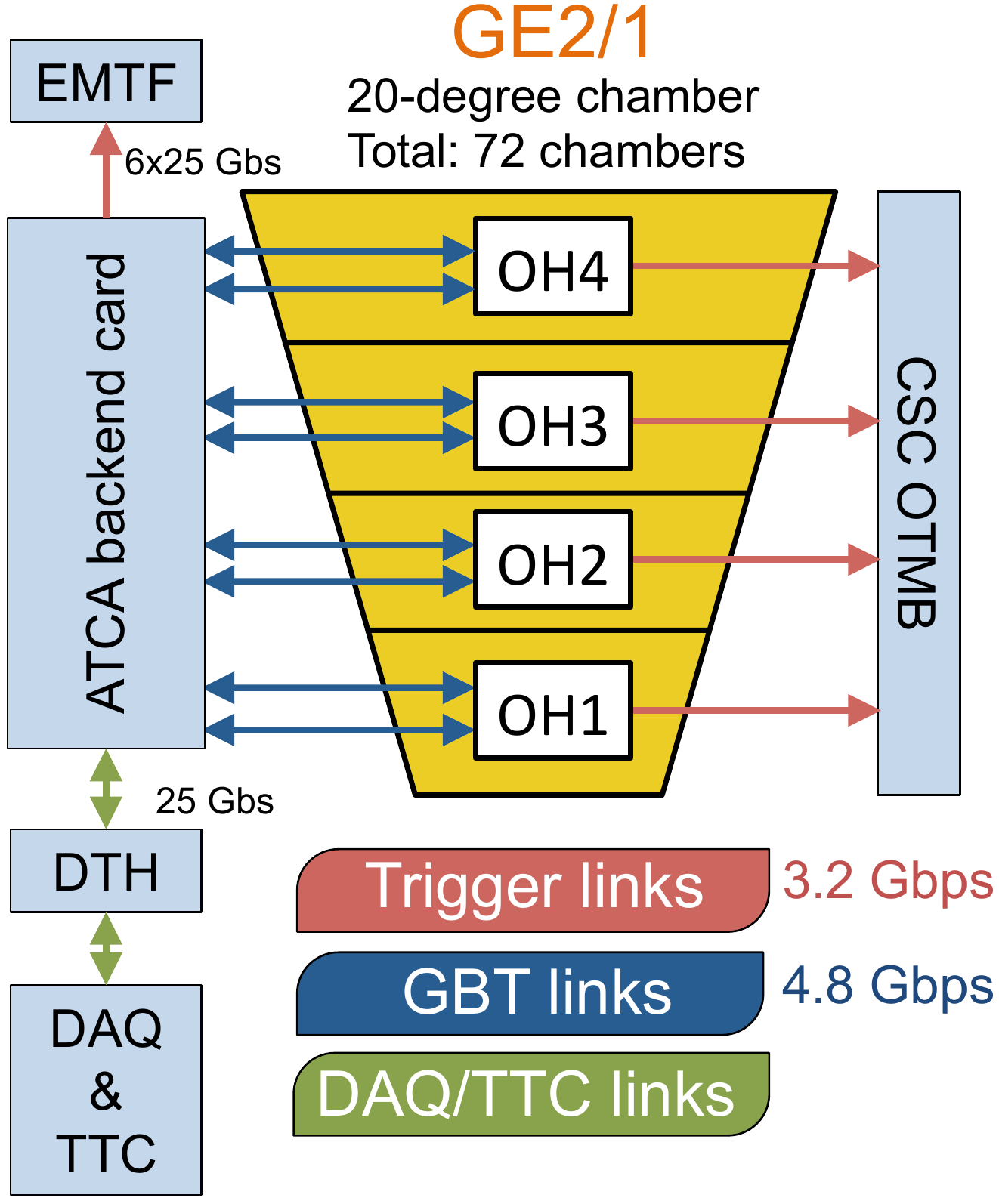}\\
\caption{\label{fig:ge21block}Full GE2/1 front chamber prototype with modules indicated (left) and block diagram of the GE2/1 electronics with frontend Opto-Hybrid (OH) boards and backend Advanced Telecommunications Computing Architecture (ATCA) boards (right).}
\end{figure}

\begin{figure}[!hbp]
\centering
\includegraphics[width=0.42\linewidth]{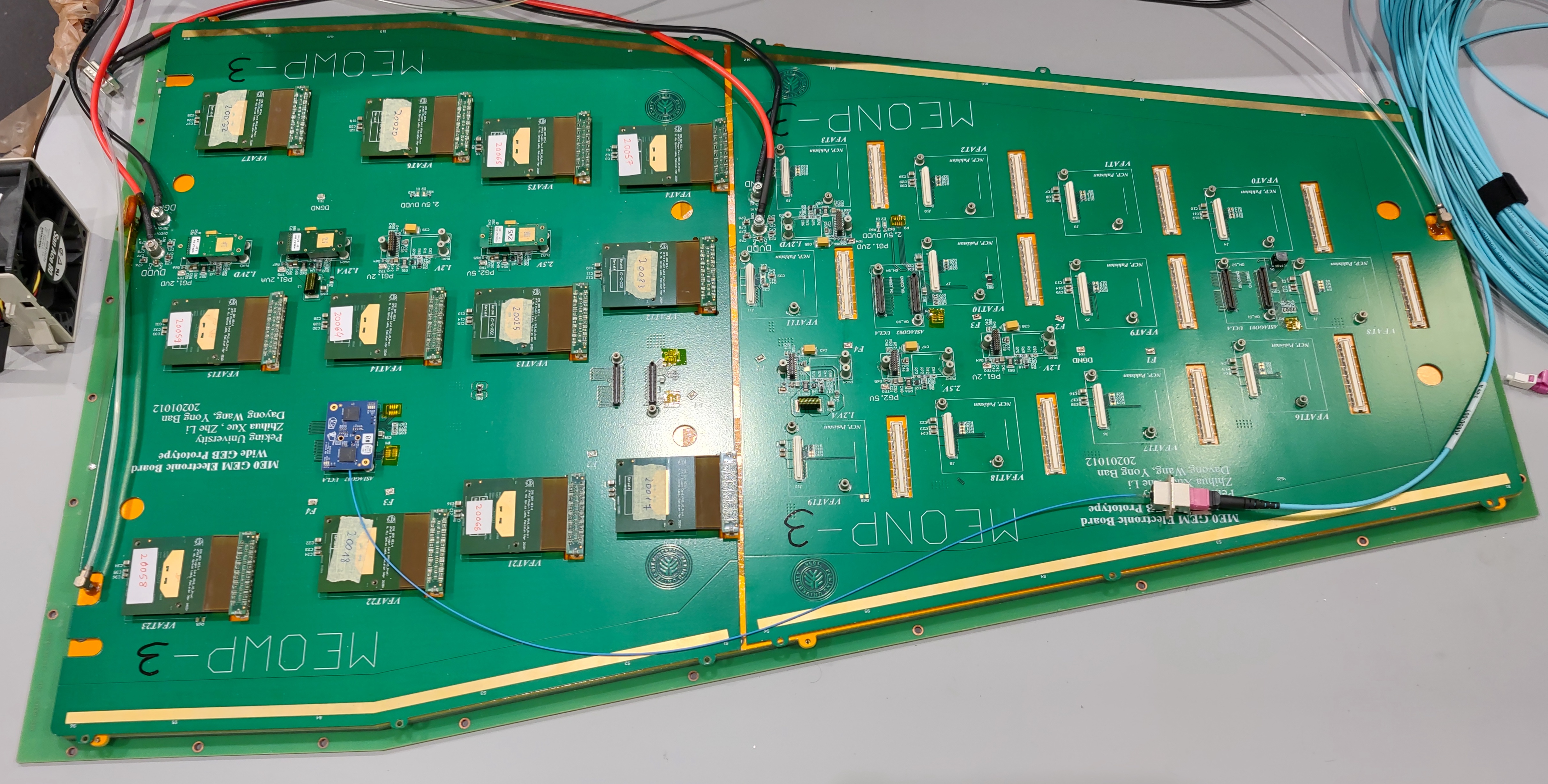}\hspace{0.2cm}\includegraphics[width=0.42\linewidth]{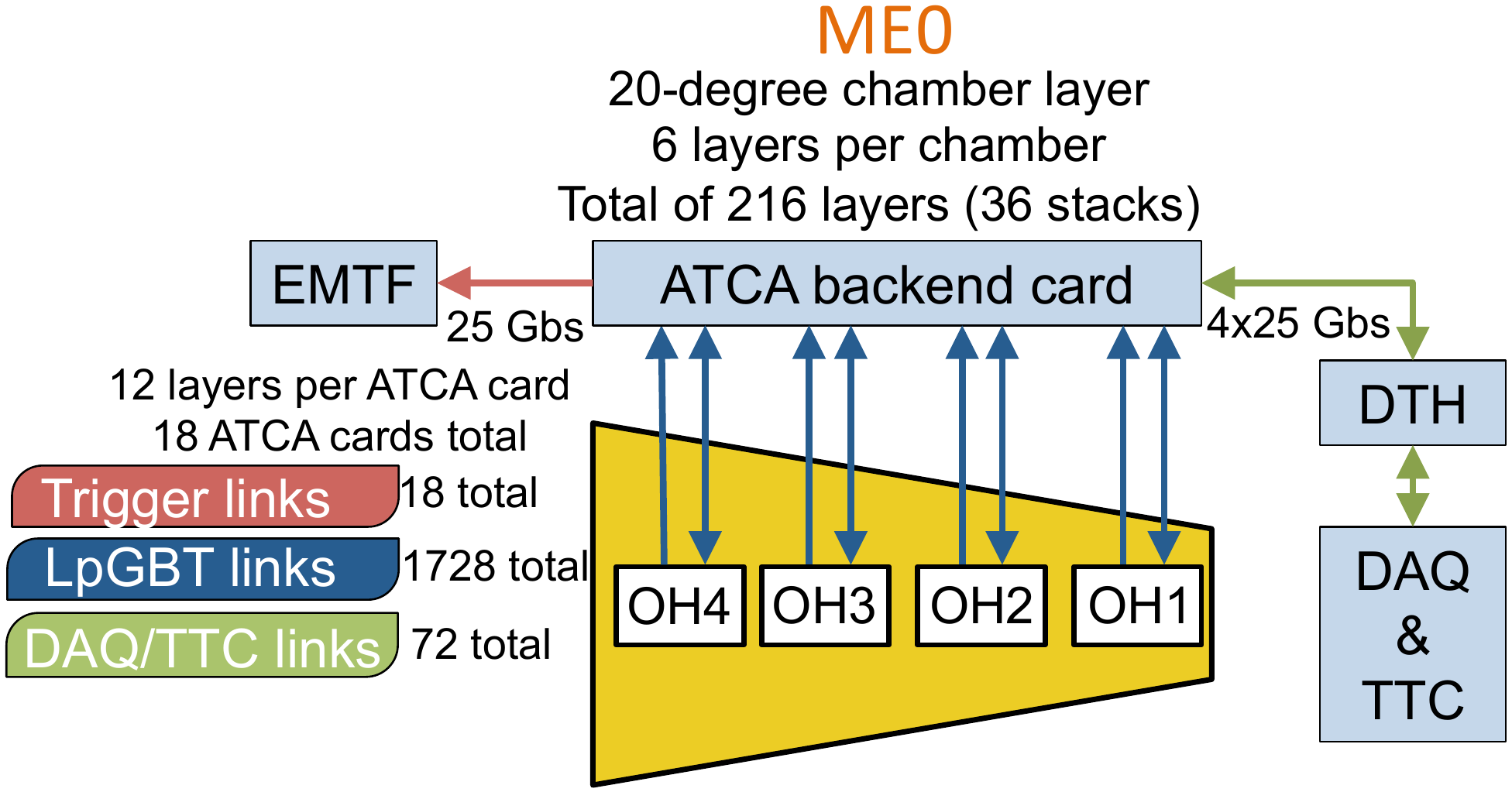}
\caption{\label{fig:me0-block}An ME0 prototype module partially instrumented with 12 frontend cards (left) and block diagram of the frontend/backend electronics for the ME0 (right).}
\end{figure}

The GE2/1 and ME0 are modular triple-GEM detector systems. Each GE2/1 chamber consists of four modules, namely the M5--M8 modules for the front GE2/1 chamber, and the M1--M4 modules for the back chamber [figure \ref{fig:ge21block} (left)]. When installed in the CMS experiment, a front and a back chamber will be mounted back-to-back, in front of the ME2/1 cathode strip chamber in the second muon station, covering the pseudorapidity range $1.62 < |\eta| < 2.43$. The ME0 detector system consists of six individual modules [see figure \ref{fig:me0-block} (left)] stacked upon one another. These ``stacks'' will be inserted in the endcap nose behind the high granularity calorimeter, increasing coverage of the muon spectrometer between $2.0 < |\eta| < 2.8$.

%\begin{figure}[!hp]
%\centering
%\includegraphics[width=0.23\linewidth]{ge21_testStand_20210811_TWEPP.jpg}\hspace{0.4cm}\includegraphics[width=0.23\linewidth]{me0_testStand_vertical}\\
%\caption{\label{fig:chambers}Full GE2/1 front chamber prototype with modules indicated (left) and block diagram of the GE2/1 electronics with frontend Opto-Hybrid (OH) boards and backend Advanced Telecommunications Computing Architecture (ATCA) boards (right).}
%\end{figure}

% ORIGINAL FIGURE BELOW:

% ORIGINAL FIGURE BELOW:
%\begin{figure}[!hbp]
%\centering
%\includegraphics[width=0.42\linewidth]{me0_horizontal}\hspace{0.2cm}\includegraphics[width=0.42\linewidth]{me0_electronicsBlockDiagram_horizontal}
%\caption{\label{fig:me0-block}An ME0 prototype module partially instrumented with 12 frontend cards (left) and block diagram of the frontend/backend electronics for the ME0 (right).}
%\end{figure}

\section{Frontend electronics for the GE2/1 and ME0}

\begin{figure}[!hbp]
\centering
\includegraphics[width=0.52\linewidth]{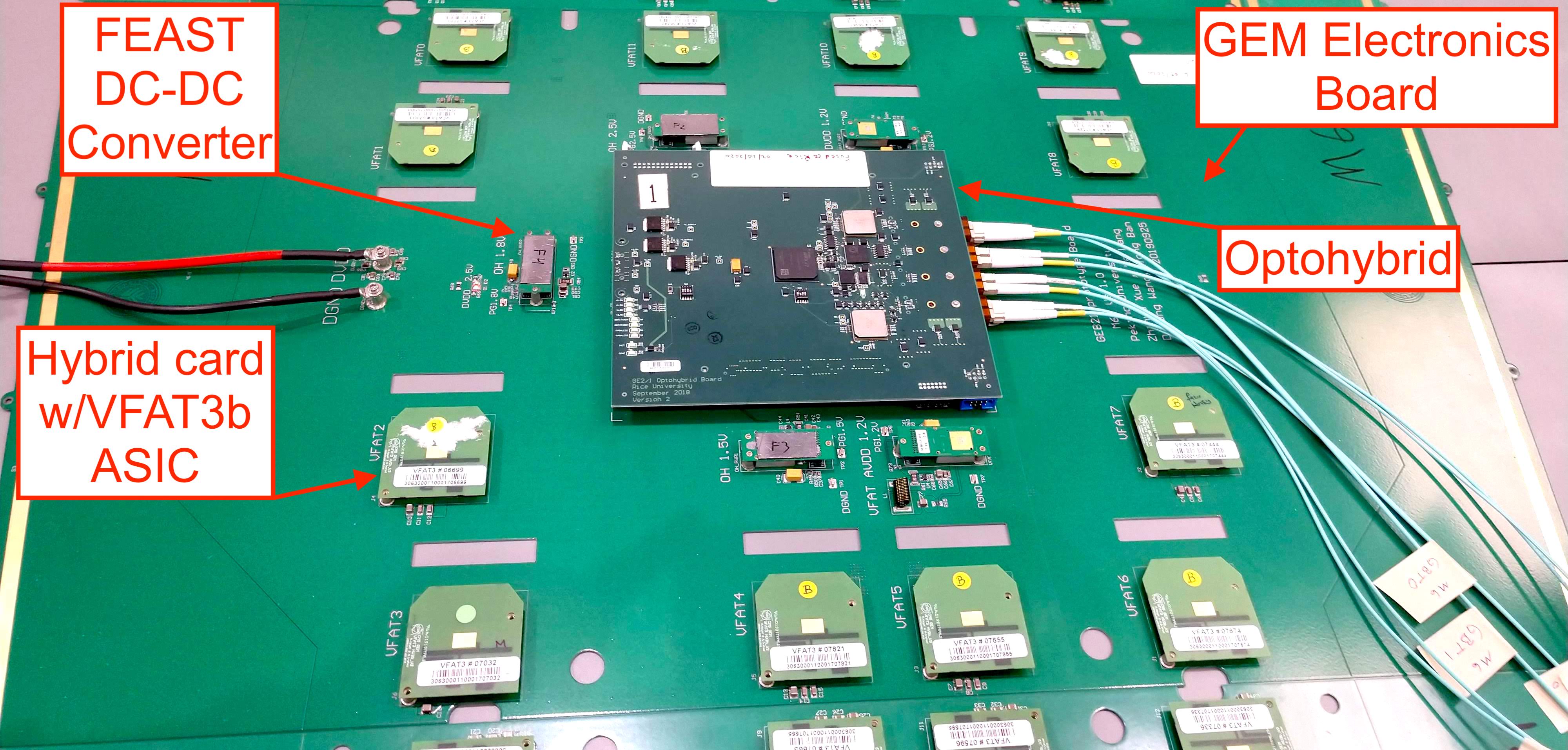}\hspace*{0.4cm}\includegraphics[width=0.19\linewidth]{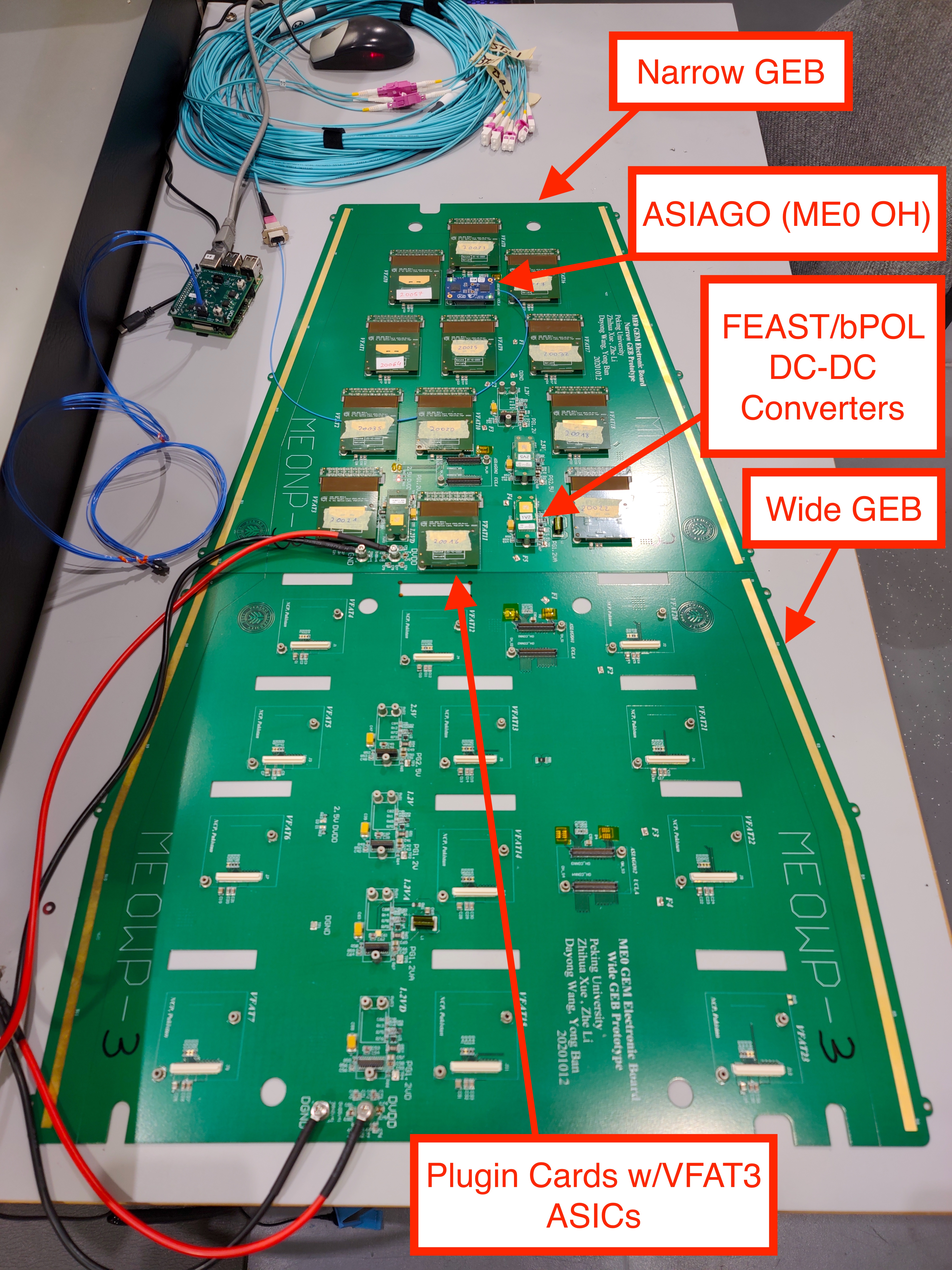}
\caption{\label{fig:testStands}A GE2/1-M6 GEB with frontend electronics indicated (left), and a set of wide and narrow GEBs for one ME0 module (right).}
\end{figure}

\begin{figure}[!htp]
     \centering
     \begin{subfigure}[t]{0.3\textwidth}
         \centering
         \includegraphics[width=\textwidth]{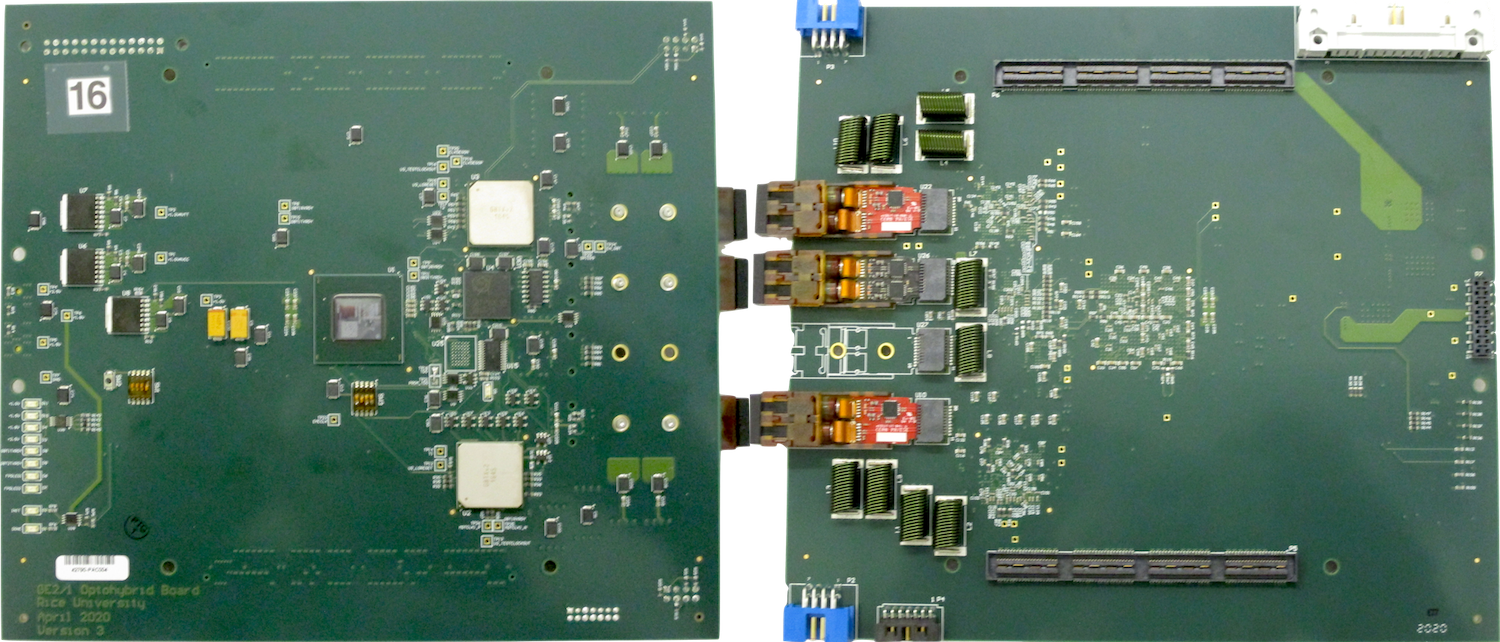}
         \caption{\label{fig:ge21-OH}GE2/1 OH front (left) and back (right) \cite{OH}.~~~~~~~~~~~~~~~~~~~~~~~~~~~~~~~}
     \end{subfigure}
     %\hfill
     \begin{subfigure}[t]{0.3\textwidth}
         \centering
         \includegraphics[width=\textwidth]{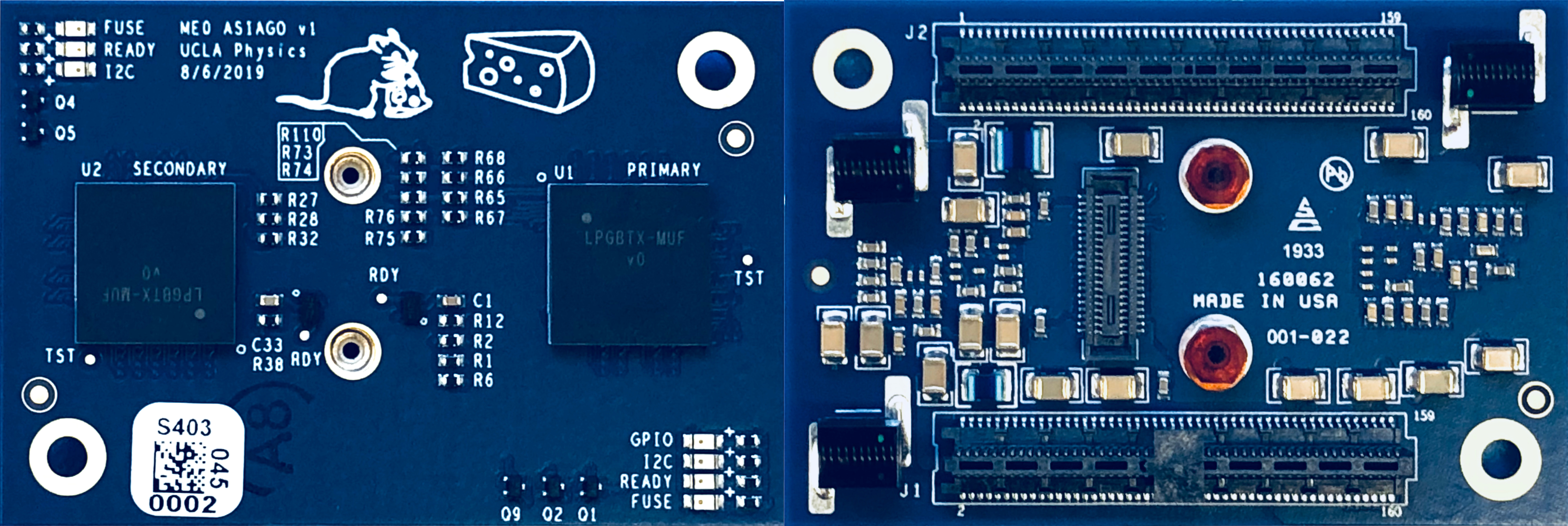}
         \caption{\label{fig:me0-OH}ME0 OH front (left) and back (right) \cite{ASIAGO}.~~~~~~~~~~~~~~~~~~~~~~~~~~~~~~~~~~~~~}
         
     \end{subfigure}
     %\hfill
     \begin{subfigure}[t]{0.3\textwidth}
         \centering
         \includegraphics[width=0.8\textwidth]{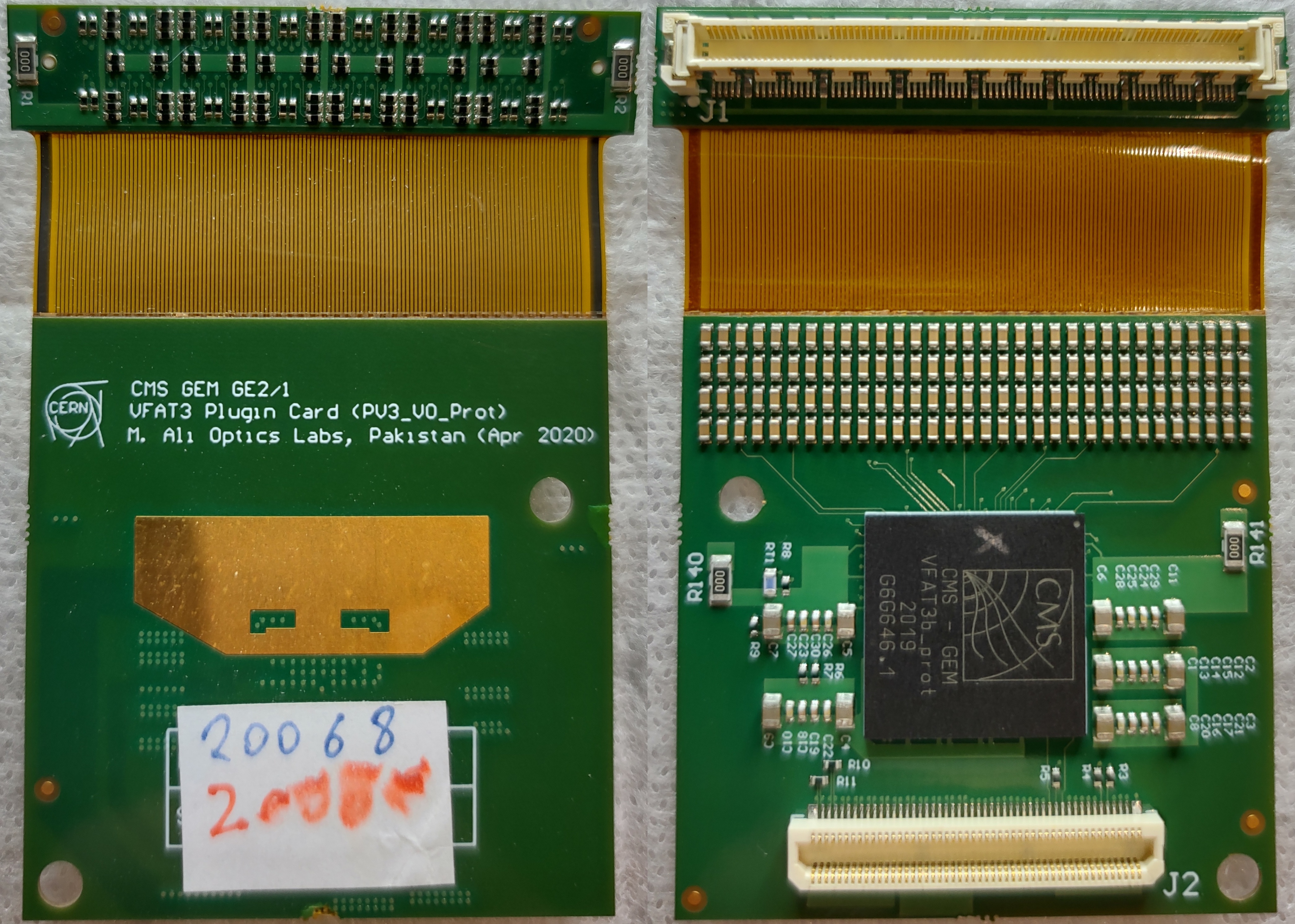}
         \caption{\label{fig:pluginCards}The front (left) and back (right) of the plugin card with VFAT3b ASIC \cite{pluginCard}.}
         
     \end{subfigure}
        \caption{\label{fig:electronics}Photos of the core frontend electronics for the GE2/1 and ME0 detector systems.}
        
\end{figure}

Both the GE2/1 and ME0 detectors share similar frontend electronics; see block diagrams for the GE2/1 and ME0 in figures \ref{fig:ge21block} and \ref{fig:me0-block} (right), respectively. For example, each GE2/1 module or ME0 layer will have a GEM electronics board (GEB), which is responsible for routing the signals from the readout (RO) application specific integrated circuits (ASICs) to the Opto-Hybrid (OH). Figure~\ref{fig:testStands} displays pictures of a GE2/1 (left) and ME0 (right) module with their electronic components indicated. The GEB is fixed to the detector readout board that the inputs of the frontend readout ASICs are plugged into. The OH is responsible for communication between the frontend and backend, transmitting both data aquisition (DAQ) and trigger data via the Versatile TransReceiver (VTRx) and Versatile Twin Transmitter (VTTx) \cite{vlProject} on the GE2/1 OH, and the VTRx+ \cite{vlPlus} on the ME0 OH. The GE2/1 OH \cite{OH} (figure \ref{fig:ge21-OH}) features a Xilinx Artix-7 field programmable gate array (FPGA), which interfaces with two gigabit transceiver (GBT) ASICs that communicate with the frontend RO chips, and the GBT-Slow Control Adapter ASIC for slow control commands. Due to the harsh radiation environment that the ME0 will operate in, the ME0 OH \cite{ASIAGO} (see figure \ref{fig:me0-OH}) features 2 low-power GBT (LpGBT) chips in lieu of an FPGA. The frontend RO ASICs are the packaged Very Forward ATLAS and TOTEM 3b (VFAT3b) chips mounted on a plug-in card \cite{pluginCard} (figure \ref{fig:pluginCards}). One GE2/1 GEB is operated with one OH, which communicates with 12 VFATs, while one ME0 GEB (two GEBs per chamber) contains two OHs, each of which communicates with six VFATs a piece. Low voltage for the GE2/1 is distributed by FEASTMP\_CLP DC-DC~converters~\cite{feast}, while the ME0 will be powered by bPOL12V DC-DC converters \cite{bPOL}. 
\section{Frontend electronics integration testing}\label{sec:integration}
Frontend electronics integration for the GE2/1 and ME0 detector systems consist of several procedures and tests to establish communication, calibrate, and determine the noise in the system. For both detectors, the integration pipeline is identical, although there are some exceptions; (e.g., the ME0 OH does not have an FPGA, thus FPGA programming is skipped during the second step of the pipeline). See table \ref{tab:FEprocedure} for a comprehensive list and explanation of the integration procedures.

\begin{table}[!htp]
\centering
\caption{\label{tab:FEprocedure}Frontend electronics testing procedure.}
%\smallskip
\resizebox{\textwidth}{!}{%
\begin{tabular}{|cc|}
\hline
\Huge{Step} 	& \Huge{Procedure}\\
\hline
\Huge{0} 		& \Huge{Test the voltages produced by the DC-DC converters}\\
\Huge{1}		& \Huge{Establish connectivity [check (Lp)GBT communication, check trigger links, program FPGA (GE2/1), GBT phase scans, check VFAT synchronization]}\\
\Huge{2}		& \Huge{Retrieve VFAT calibration data from the database and program nominal register values}\\
\Huge{3}		& \Huge{Calibrate VFATs by performing digital-to-analog converter scans to characterize and determine nominal values of programmable registers}\\
\Huge{4}		& \Huge{Equivalent noise charge measurement for DAQ electronic links}\\
\Huge{5}		& \Huge{S-bit noise measurement for trigger eLinks}\\
\hline
\end{tabular}%
}
\end{table}

Connectivity testing begins with first establishing communication with the OH, checking the trigger links, (programming the FPGA for GE2/1), and scanning the clock phases of the (Lp)GBTs to synchronize the VFATs to a common clock.

%Example output in figure \ref{fig:phase-and-dacScans} (left) shows successful results for ME0 phase scans.

%\begin{figure}[!htp]
%\centering
%\includegraphics[width=0.3\columnwidth]{phaseScans_20210902_1V3FEASTs}
%\caption{\label{fig:phaseScans}Phase scan results for the second OH on the wide ME0 GEB.}
%\end{figure}

Next, calibration data (internal current/voltage biases) are retrieved from the CMS GEM database, which are subsequently programmed into the VFATs. Characterization scans for each of the programmable digital-to-analog converters (DACs) on each VFAT are performed by scanning a range of register values and reading the output current/voltage with the internal analog-to-digital converter (ADC) of the VFAT. Optimal DAC values are determined from the resulting curves.
%see an example for the second constant fraction discriminator (CFD) below in figure \ref{fig:phase-and-dacScans} (right) on a GE2/1-M7 module. %This step ensures that each VFAT is properly calibrated, thus returning accurate data.

\begin{figure}[!htp]
\centering
\includegraphics[width=0.26\linewidth]{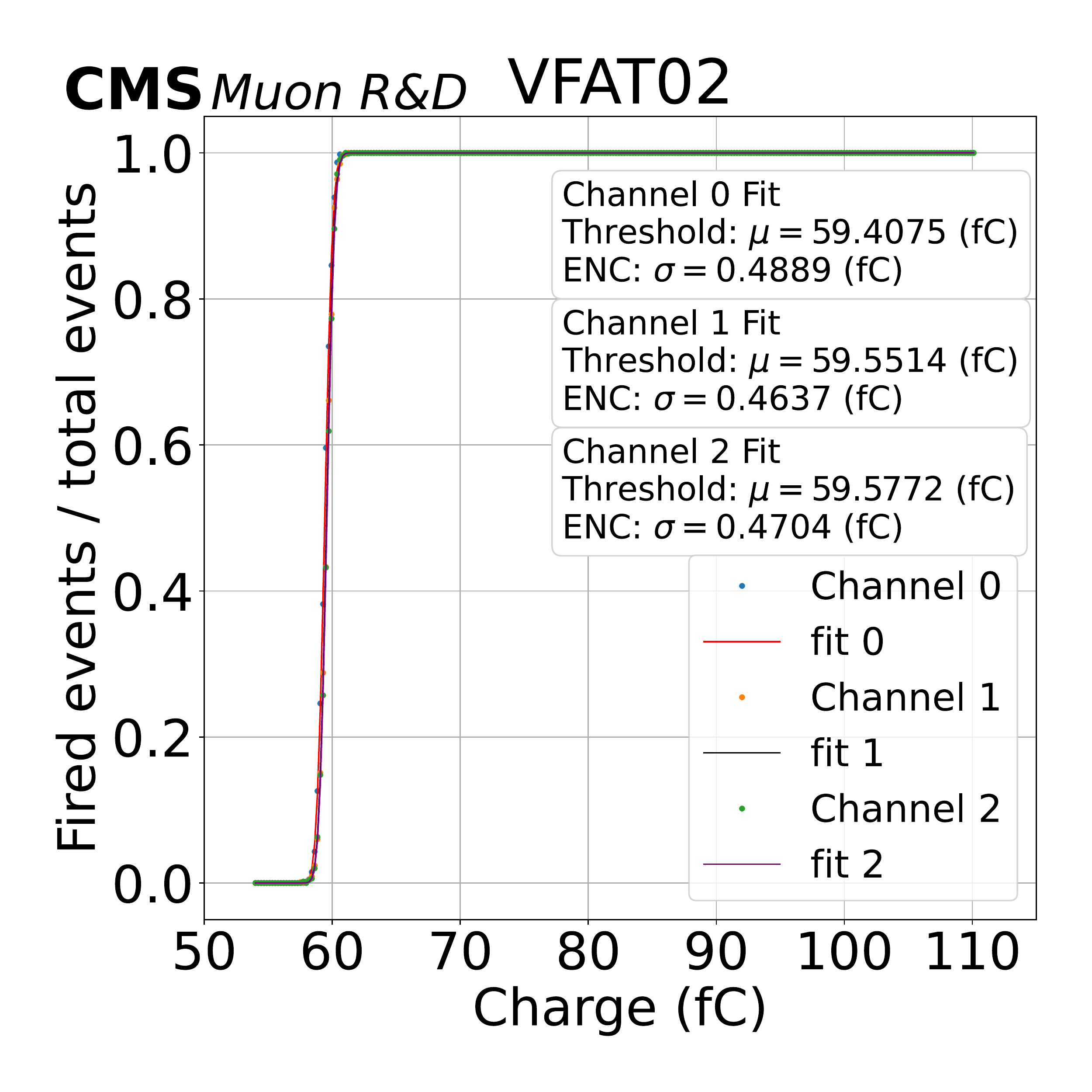}\hspace{0.2cm}\includegraphics[width=0.26\linewidth]{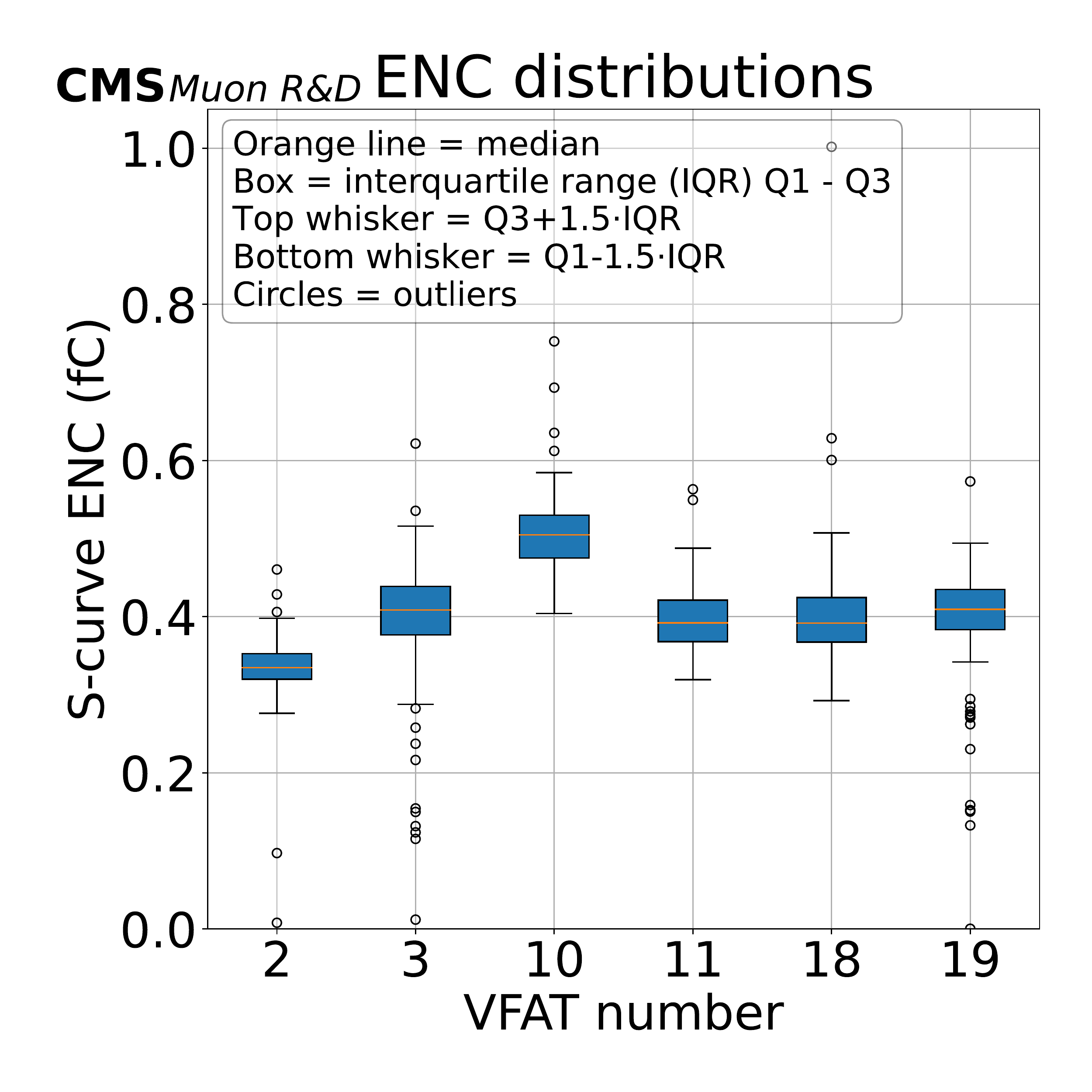}
\caption{\label{fig:ENCandDist}S-curves for three channels with fit data (left). S-curve sigma (ENC) distributions for all 128 channels for six VFATs on an ME0 GEB (right). Note that one outlier point for VFAT18 is not displayed.}
\end{figure}

\begin{figure}[!hp]
\centering
\includegraphics[width=0.27\linewidth]{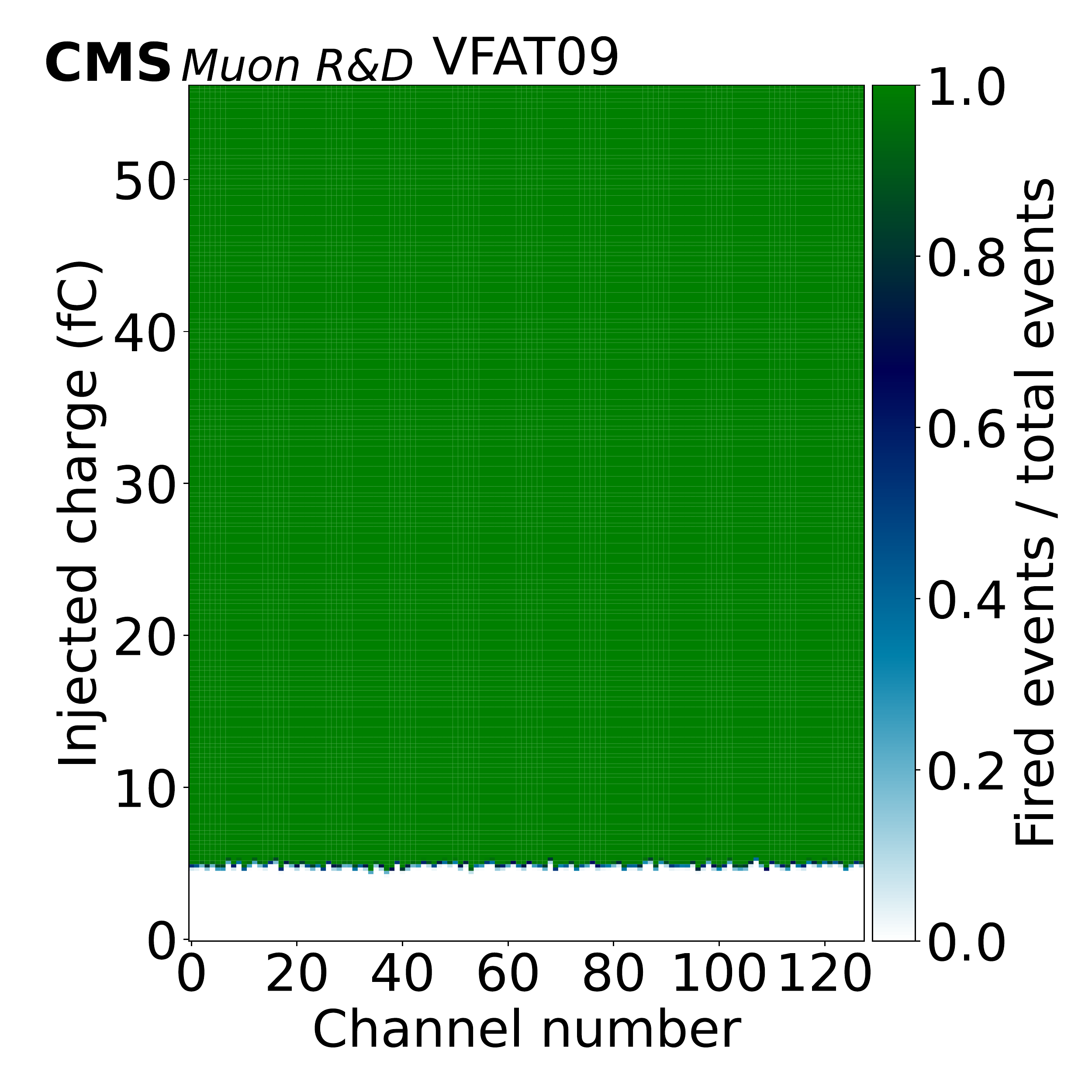}\hspace*{0.2cm}\includegraphics[width=0.29\linewidth]{vfat17_sbit04_TWEPP_GMMapproved_sbit_rate.pdf}
\caption{\label{fig:2Dhist_sbit}S-curves for all channels (left) and trigger S-bit noise rate measurement (right) for a single S-bit on one VFAT on an ME0 detector.}
\end{figure}

%\begin{figure}[!htp]
%\centering
%\includegraphics[align=c,width=0.42\columnwidth]{phaseScans_20210902_1V3FEASTs}\hspace*{0.2cm}\includegraphics[align=c,width=0.24\columnwidth]{DAC_scan_M7-P4_v10_GMMapproved}
%\caption{\label{fig:phase-and-dacScans}Successful phase scan results for the second OH on the wide ME0 GEB (left) and DAC characterization scan for the second constant fraction discriminator on a VFAT on a GE2/1-M7 module (right).}
%\end{figure}
Equivalent noise charge (ENC) measurements (colloquially referred to as ``S-curves'' due to their shape) are then taken for the DAQ electronic links (eLinks) by injecting increasing amounts of charge into each channel and recording the comparator response, and then fitting a modified error function [eq.~(\ref{eqn:scurves})] to the data:
\begin{equation}\label{eqn:scurves}\tag{1}
f(q) =A\cdot\mathrm{erf}\bigg(\dfrac{\mathrm{max}(p,q) - \mu}{\sigma\sqrt{2}}\bigg) + A
\end{equation}
where $q$ is the magnitude of injected charge, $A=n/2$, where $n$ is the number of injected charges, $p$ is the pedestal (recorded comparator responses with no injected charge), $\mu$ is the comparator threshold, and $\sigma$ is the ENC derived from the fit. S-curve results for a few channels are shown in figure \ref{fig:ENCandDist} (left); ENC distributions for six VFATs are shown in figure \ref{fig:ENCandDist} (right), and S-curves for all 128 channels on one VFAT are shown in figure \ref{fig:2Dhist_sbit} (left), for an ME0 detector. 
The last step is the noise measurement for the trigger eLinks. For triggering purposes, the RO strips (connected to the VFATs) are paired, resulting in 64 strip groups, or ``S-bits.'' For GE2/1, the RO strips in the vertically adjacent or neighboring eta sectors are paired (i.e., channel 1 from one VFAT is paired with channel 1 from another VFAT in the neighboring RO sector), while in ME0, horizontally adjacent strips in the same RO sector are paired together. When one (or both) of the two strips is (are) triggered, an S-bit is recorded. To determine the noise, the threshold is fixed, and one by one each S-bit is unmasked and counts are recorded. This is repeated for the entire DAC range of $[0,255]$. The noise is thus determined by considering each individual S-bit [see an example in figure \ref{fig:2Dhist_sbit} (right)], and the cumulative sum and average of all S-bits for each VFAT.

\section{Summary of GE2/1 and ME0 electronics integration status and plans}
With the GE2/1 prototype electronics integration efforts at CERN, Rice University, Texas A\&M, and Florida Tech coming to a close, mass production of GE2/1 GEM detectors has been initiated. All electronic components for GE2/1 are projected to be manufactured and tested by February of 2022, and full electronics integration with manufactured chambers expected to be complete by February of 2023. The ME0 electronics integration effort is ongoing, with component testing and electronics integration at CERN, UCLA, and Florida Tech. Recently, the second version pre-production OH was approved for manufacturing, and will begin production once the version 1 LpGBTs have been tested. The ME0 GEB is in its final prototyping stages, which will soon be reviewed for production. The collaboration is also currently testing an ATCA backend card with various FPGAs and optics, with a final design to be ready by early 2022. Manufacturing and testing of the ME0 on-chamber electronics is projected to be completed by November 2022, with all detector stacks to be installed in the new endcap nose by February 2026. 

\section*{Acknowledgments}
We would like to thank the U.S. DOE (HEP) and the NSF (Cornell University subaward, P.I. Dr. Marcus Hohlmann) for their support of this work, and all members of the CMS GEM group for their contributions to this project.

%\bibliography{bibfile} % a bibtex command

%\printbibliography   % a biblatex command

\end{document}